**Title: From a sounding rocket per year to an observatory per lifetime**

Author: Martin C. Weisskopf [1]

Date: March 15, 2013

Abstract

I summarize the excitement of my role primarily in the early years of X-ray Astronomy. As a "second-generation" X-ray astronomer, I was privileged to participate in the enormous advance of the field, both technically and astrophysically, that occurred in the late 1960's and 1970's. The remainder of my career has concentrated on the design, construction, calibration, operation, and scientific maintenance of the "cathedral" that is the Chandra X-Ray Observatory. I contrast my early experiences with the current environment for the design and development of instrumentation—especially X-ray optics, which are absolutely essential for the development of the discipline. I express my concerns for the future of X-ray astronomy and offer specific suggestions that I hope will advance the discipline at a more effective and rapid pace.

The Columbia Years

I spent my early post-graduate years from 1969 until the fall of 1977 as part of Robert (Bob) Novick's group at the Columbia Astrophysics Laboratory (CAL), Columbia University in the city of New York. To say that these were exciting and interesting times would be an understatement. For example, a few days after I arrived, I found myself in the entrance hall of the Pupin building sitting with T. D. Lee and a number of other faculty members holding long discussions with (mostly) student demonstrators that were attempting to take over this physics building in protest over Columbia scientists' participation in a military think tank.

Even before I accepted the position at CAL, Bob took me to a meeting at the company American Science and Engineering (AS&E), in Cambridge, Massachusetts. The discussion centered on something called a "Super Explorer" and the "Principal Investigator Group" (acronyms withheld by popular request). At this meeting, I first met several people—especially, Leon Van Speybroeck and Harvey Tananbaum—who were to play significant roles in my career. Riccardo Giacconi made the most profound impression: He emphasized the importance of imaging for advancing X-ray astronomy, an insight he formed within a year after the 1962 sounding-rocket experiment that discovered the brightest non-solar source, Scorpius X-1. It is little wonder he eventually (2002) received the Nobel Prize.

It puzzles me even to this day that, despite the leaps and bounds made by ever improving angular resolution, there are many proponents seeking more photons (larger area) at the price of angular resolution. I don't imply that such experiments haven't proven and won't prove fruitful, but frankly I feel they pale in comparison to what has been, could be, and should be accomplished.

---

[1] Space Sciences Office, NASA/Marshall Space Flight Center, Huntsville AL 35812

Just after I arrived at the CAL, a rocket that Bob Novick and those already there had put together blew up on launch at the White Sands Missile Range, New Mexico. The payload featured the first X-ray "telescope", a concentrator, made of dozens and dozens of gold-coated microscope slides mounted to approximate a paraboloid.  I mention this because those circumstances taught me (and the rest of us) an important lesson that influenced my approach to interacting with NASA and especially my approach to Chandra. The reason that the rocket blew up was that the second-stage liquid-fuel Aerobee never ignited. The liquid-fuel stage stood on a metallic "milk stool" above the solid-rocket Nike booster that started the journey into space.  As the Nike's acceleration built up it passed through the milk stool and entered the Aerobee's liquid fuel tanks. At least the event was spectacular. The firing of the liquid fuel was to have been triggered by means of a lanyard connecting the ignition system to the launch tower; but someone forgot to connect the lanyard. Of course there were investigations and blame and this was not the scientist's responsibility, but to us the message was clear: We hadn't paid enough attention to everything involved in the entire system. When the rocket failed, ultimately we scientists pay the biggest price.

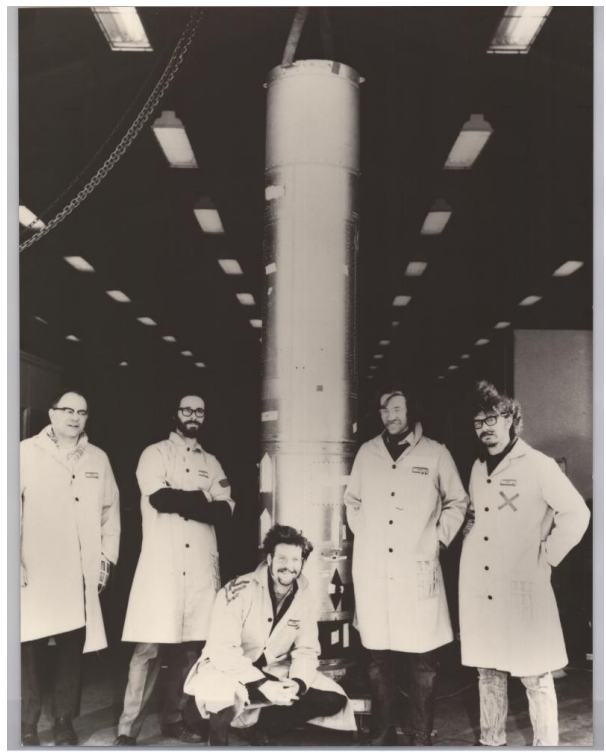

*Figure 1. Picture taken in 1970 at Wallops Island, of sounding rocket 17.09, which featured two types of X-ray polarimeters. This experiment unambiguously measured the polarization of the integrated X-ray emission from the Crab Nebula. From left to right: Robert Novick (CAL Director), Gabriel Epstein (my graduate student), myself, Richard Wolff (my office mate), and Richard Linke (his graduate student).*

During these early times, there were excellent groups forming at several US institutions: Columbia, Massachusetts Institute of Technology (MIT), AS&E, California Institute of Technology (CIT), Wisconsin, Naval Research Laboratory (NRL), Lawrence Berkeley Laboratory (LBL), Goddard Space Flight Center (GSFC), Lockheed Palo Alto, Stanford, etc.). All were more or less adequately funded to develop and demonstrate instrumentation (including X-ray optics) capable of obtaining scientific results. In the early 1970's, these capabilities were demonstrated mainly through sounding-rocket experiments. At Columbia we utilized sounding rockets with a passion. With four young Assistant Professors (Paul Van den Bout, Roger Angel, Richard Wolff, and myself) supplementing Bob's fertile imagination and abilities as a hands-on experimentalist, we developed spectrometers, concentrators, telescopes, and polarimeters for X-ray astronomy. We were designing, building, and flying new instruments at a cadence of about once per year. Like most of the other groups, we were all (to a greater or lesser degree) participating in every experimental aspect for doing the science of X-ray astronomy, not necessarily specializing in one arena or another. The competition and rivalry (for the most part friendly) was an essential ingredient in developing the field. The best technical approaches often resulted from a merging of ideas, techniques, and approaches from different organizations.

During the 1970's at CAL, we also began participating in the satellite era—successfully proposing, building, testing, and flying both an X-ray spectrometer (solar and stellar) and an X-ray polarimeter (stellar) on the OSO-8 satellite. Our pioneering efforts to establish the field of X-ray polarimetry were sadly the precursor to a frustrating future. Based upon early successes, Bob led the development of a polarimeter for the first (Russian) Spectrum-X satellite mission, which unfortunately was ultimately cancelled. More recently, NASA cancelled a GSFC-led Small Explorer Mission dedicated to low-energy X-ray polarimetry.

Simultaneously, CAL partnered with a number of institutions to develop instrumentation for the High-Energy Astronomy Observatory (HEAO) series of satellites. Our team at Columbia had a major role in three of the originally four HEAO satellites, including a one-arc minute, 1000-cm$^2$ Kirkpatrick-Baez telescope in collaboration primarily with AS&E/SAO (Bob Novick and Paul Gorenstein running the show). This experiment would have performed an all-sky survey on the first HEAO mission. Unfortunately the HEAO program hit a major obstacle, such that I (along with the rest of the community) experienced the first of a series of political decisions that seemed to relegate science second to presumed expediency. Frankly, the cancellation and resurrection of the HEAO program came as quite a shock to this naïve young researcher. Clearly the HEAO Program was excellent, with cutting-edge science and technology: Why should it be cancelled? Obviously I had a lot to learn about NASA politics. However, the experience was, in its own way, invaluable. Long story short, the HEAO program with its 4 satellites was cancelled. The new and reduced program resurrected from its ashes sadly did not completely encompass the best science. Historians will tell us that the decisions were politically necessary. Perhaps, but one is never sure. Can you imagine the progress of X-ray astronomy had the arc-minute-resolution all-sky survey, which was the first of the original HEAO series, been performed in the 1970's?

# The Marshall Years

In 1977 I received an offer from NASA to go to the Marshall Space Flight Center (MSFC; Huntsville, Alabama) to become the Project Scientist for the Advanced X-ray Astrophysics Facility (AXAF), eventually renamed the Chandra X-ray Observatory. I still hold this position. MSFC, in partnership with Riccardo Giacconi and scientists at the Smithsonian Astrophysical Observatory (SAO), had won management of this potential mission in competition with the Jet Propulsion Laboratory (JPL)/CIT and with GSFC. The community wanted to avoid many of the difficulties encountered in accomplishing the HEAO-2 (*Einstein* Observatory) mission: A critical element was to have Project Science on-site rather than long-distance, as was the case for the Einstein Observatory. At the risk of appearing immodest, I am firmly convinced that this decision was a major factor in Chandra's programmatic, technical, and scientific success. The second critical element was that the Project-Science function was not to be implemented by a single person but rather by a team of scientists at MSFC and at SAO, which brought all the *Einstein* experience to bear on this challenging project. Chandra was to be the mission that, amongst many other objectives, would address questions raised by the diffuse glow of X-rays also detected during that first sounding-rocket experiment in 1962. To accomplish the task of resolving the "diffuse" X-ray background would require an angular resolution of an arcsecond or better and an effective area of several hundred square centimeters. Although no one had ever built such an X-ray telescope, the scientific requirements were clear. Moreover, the community united to try to hold the line on requirements, placing scientific—not programmatic—considerations at the forefront. We were reasonably successful in this, primarily for the telescope. (As noted below, this is neither the time nor the place for my unabridged version of the Chandra saga.) In 1977, the projected launch of Chandra was 1985. For a number of reasons, both technical and programmatic (mostly financial), the projected launch date slipped a year per year until 1992. With consistent funding and only minor delays thereafter, the launch occurred in 1999.

The technical insights and experience provided by Chandra Project Science (not just me, but also Ron Elsner, Allyn Tennant, Brian Ramsey, Steve O'Dell, Marshall Joy, Jeff Kolodziejczak, Doug Swartz, and several others who were part of the Project Science Team at MSFC) and by SAO scientists and engineers, along with all of our sense of responsibility to the rest of the community, served the project in numerous ways. Perhaps the most important was the control of requirements, which had no major change over the 22 years between 1977 and the 1999 launch. Our ability to accomplish Chandra with such success was founded on the experiences and capabilities that we developed in the "early days". We had built instruments, sometimes making last minute repairs while the rocket was mounted in the launch tower; we had participated in satellite missions, sometimes going through the agonies of cancellation, etc. In other words, by and large, we knew what we were doing!

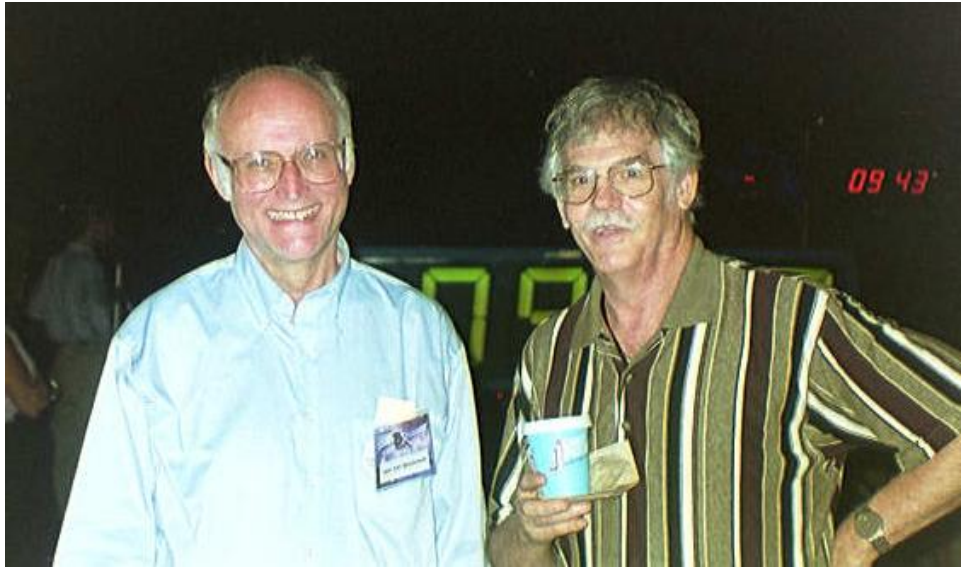

*Figure 2. Leon Van Speybroeck (Chandra Telescope Scientist) and myself during a hold 9 minutes 43 seconds before the first launch attempt. The third attempt on 1999 July 23 succeeded!*

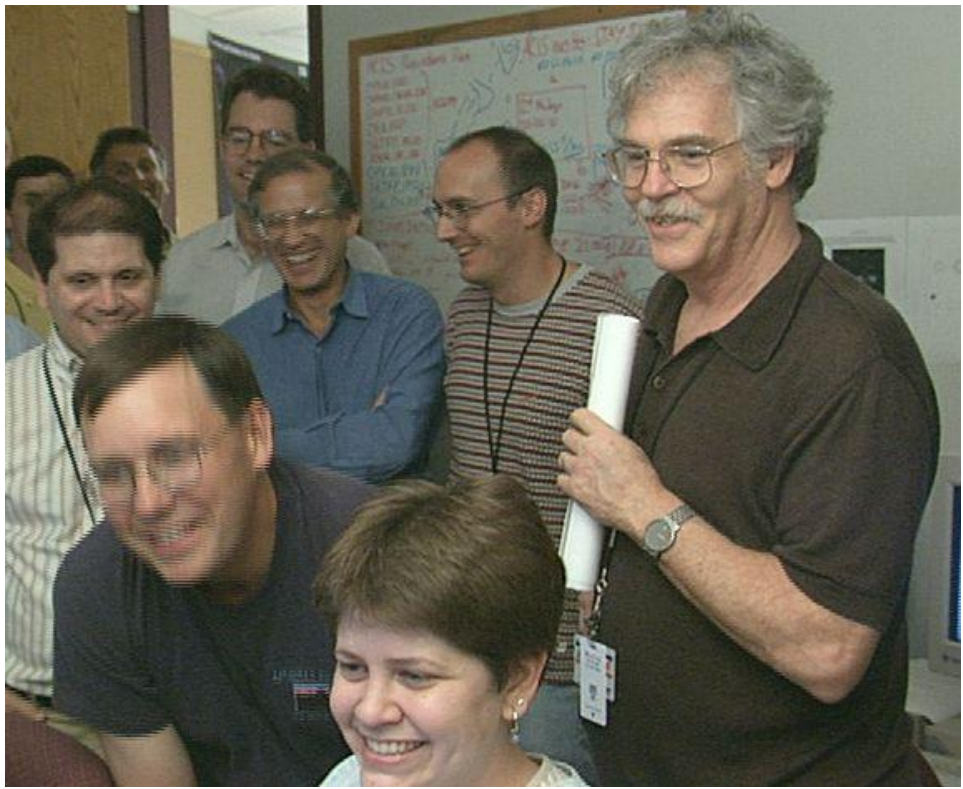

*Figure 3:* *From right to left: myself, Tom Aldcroft, Catherine Grant, Harvey Tananbaum, Roger Brissenden, Mark Bautz, Mark Freeman, Fred Baganoff, and Ken Gage at the Chandra Control Center, sharing the excitement of the official first-light observation in 1999 August.*

As noted above, this is neither the place nor the time for me to write my version of Chandra's history and accomplishments, if for no other reason than that the Observatory—originally designed with a formal lifetime requirement of 3 years and a goal of 5—is still obtaining outstanding scientific results. Indeed, at this writing (2013 March 5), the most famous supernova remnant, the Crab Nebula, is flaring in γ-rays: Chandra will be pointed at the Crab this evening, participating in the important hunt for the specific location within the Nebula of the γ-rays flares.

## The Future

I want to conclude this brief narrative with some comments and personal insights for the future of X-ray astronomy. First, I am (and many of my colleagues are) very concerned over the ever shrinking number of organizations (especially academic) that are currently directly involved in building scientific hardware. If the current trend continues, there may eventually be only one US institution with the infrastructure and resources to accomplish such tasks: That institution will most likely be at one of the NASA centers. Having fewer and fewer institutions responsible for technology development is not only worrisome, but potentially wasteful. Indeed, I see it as a hindrance to efficient progress. The concise and oversimplified reasons are (1) that no one has all the answers and (2) that competition breeds innovation.

How have we gotten into this situation? There are many answers and I would not presume to be able accurately to cite them all. One major factor is that we have moved X-ray astronomy into the observatory era: Chandra is a prime (and highly successful) example. Chandra has spawned the growth of a large general-observer community. These observers are vital to the discipline, not only for their scientific acumen, but also because they comprise a large (and vocal) advocacy community. Without their widespread support, it becomes harder to raise money etc.  At the same time, however, these great observatories have taken decades to build and cannot adequately serve as training grounds and development programs for optics and instrumentation that will meet the future needs of the discipline.

I believe that we pay another, more dreadful price, for the lack of young active experimentalists. As our scientific requirements grow in scope, projected costs are no longer based upon accomplishments or even partial accomplishments. We must predict years, even decades, in advance how much a new mission will cost. As the technology hasn't been demonstrated (perhaps not even conceived), the cost estimates become outrageously high. This of course delays missions, often leading to programmatic rather than science-driven technical decisions because one is trying to cost to a schedule that is totally unrealistic. We, as a community, must shoulder much of the blame. We cannot seem to stand fast in supporting a more rational approach for developing missions. Even in those cases where future requirements are (somewhat oversimplifying) crystal clear (at least to me) and currently unobtainable, the Chandra example is an excellent one to follow. Thus, to be able to detect fluxes from galaxies at the dawn of the early universe will require sub-arcsecond optics of collecting power easily a factor of 10 or more larger that Chandra. Thus, we already know what we want to, even must, do to advance the field.

To me this situation is analogous to the situation at the conception of Chandra. The telescope then was beyond the current state of the art and *competing* technologies were investigated both by the NASA Project and its research partner at SAO, but also with industry. We had not the chutzpah to put forth seriously the mission for consideration until we had built and X-ray tested an X-ray optic that met all of the detailed Chandra requirements. (This latter resulted in part because NASA and Congress insisted!) How successful was this approach? The cost overrun at launch was less than a few percent and the final cost was, accounting for inflation, the same as had been estimated by the Project and provided as input to the National Academy of Sciences Decadal Survey for Astronomy and Astrophysics for the 1980's—i.e. nearly 20 years before it was built and launched! Once again, I maintain that one of the principal reasons for this success was the heavy involvement of experimental X-ray astronomers in all phases of the program. The question then becomes, where are these experimentalists to come from now?

Of course, the answer to this question is to expand technical research primarily in the arena of X-ray optics and to enlarge the balloon and sounding-rocket program. This is nothing new: Indeed, Decadal Surveys and advisory committees have been advocating these ideas for years. However, the missing ingredient is money. I am not so naïve to suppose that NASA can significantly expand the balloon and sounding-rocket programs through an increase to any current NASA budget. The budgetary trend, if anything, is in the opposite direction. My suggestion is to set aside a non-trivial amount $25M-$40M per year from the Explorer budget to add to the existing funds already budgeted for these methods of providing rapid access to space. I realize that there may be legal difficulties associated with the Congressional language for the Explorer program, but I am confident that, if this is what the community demands, it can happen.